\documentclass[aps,prd,twocolumn,superscriptaddress,nofootinbib]{revtex4-2}
\usepackage[colorlinks=true,citecolor=red,urlcolor=blue]{hyperref}
\usepackage{amssymb}
\usepackage{amsmath,color}
\usepackage{graphicx}
\usepackage{bbm}
\usepackage{verbatim}
\usepackage[T1]{fontenc}
\usepackage[utf8]{inputenc}
\usepackage[normalem]{ulem}

\usepackage{url}
\usepackage{xcolor}
\usepackage{caption}
\usepackage{subcaption}

\usepackage{diagbox}
\usepackage{color}
\usepackage{multirow}
\usepackage{hhline}
\usepackage{tabularx}
\usepackage{orcidlink}

\begin{document}

%\preprint{}

\title{Precision Joint Constraints on Cosmology and Gravity Using Strongly Lensed Gravitational Wave Populations}


\author{Xinguang Ying\orcidlink{0009-0000-6778-2228}}
%\email[]{}
%\thanks{}
%\affiliation{}
\affiliation{School of Physics and Technology, Wuhan University, Wuhan 430072, China}

\author{Tao Yang\orcidlink{0000-0002-2161-0495}}
\email[Corresponding author: ]{yangtao@whu.edu.cn}
%\thanks{}
%\affiliation{}
\affiliation{School of Physics and Technology, Wuhan University, Wuhan 430072, China}

\date{\today}

\begin{abstract}
We present a Bayesian framework  to jointly constrain the Hubble constant ($H_0$) and the post-Newtonian parameter ($\gamma$), a key indicator of deviations from general relativity, using the population characteristics of strongly lensed gravitational wave (GW) events from binary black hole mergers. Our method extracts cosmological and gravitational information directly from the statistical properties of lensed GW populations, without relying on electromagnetic counterparts of the GW events, waveform modeling, and resolved stellar kinematics of the lens galaxy. This establishes lensed GW statistics as a clean and independent probe of cosmic expansion and gravitational physics. Assuming a flat $\Lambda$CDM cosmology and simulating a GW population observed by the third-generation detector Einstein Telescope, we demonstrate that this method can achieve precision levels of $0.60\%\sim0.99\%$ for $H_0$ and $0.53\%\sim3.3\%$ for $\gamma$ with various priors of matter density,  significantly outperforming existing joint constraints, which typically achieve $2\%$ precision on $H_0$ and $20\%$ precision on $\gamma$. These results highlight the potential of lensed GW population statistics as a robust and efficient tool for probing both the expansion history of the Universe and the nature of gravity.
\end{abstract}

\maketitle

\section{Introduction}

The first direct detection of gravitational waves (GWs) by LIGO/Virgo \cite{LIGOScientific:2016aoc} marked the beginning of a new era in gravitational wave astronomy.
Since the first observation of the binary black hole (BBH) merger event GW150914, an expanding catalog of gravitational wave (GW) events has been compiled \cite{LIGOScientific:2018mvr, Zackay:2019tzo, Venumadhav:2019lyq, Nitz:2020oeq, LIGOScientific:2020ibl, KAGRA:2021vkt,LIGOScientific:2025slb}, enabling a wide range of astrophysical and cosmological applications (primarily for the Hubble constant). Notably, GWs can act as ``standard sirens'', providing a novel method for measuring cosmological parameters \cite{Schutz:1986gp, Holz:2005df, Nissanke:2013fka}, independent of traditional electromagnetic(EM)-based approaches, and offer a promising approach to addressing the Hubble tension -- a nearly $5\sigma$ tension between the value of the Hubble constant measured by the SH0ES project in the late Universe \cite{Riess:2021jrx} and that inferred from the \textit{Planck} cosmic microwave background \cite{Planck:2018vyg}.

%free from the systematic error of the traditional electromagnetic (EM) approaches, such as the dust extinction. 
%The luminosity distance can be directly inferred from the GW waveform, and when combined with redshift information, obtained either through EM counterparts or host galaxies, it allows constraints on cosmological parameters such as the Hubble constant $H_0$ and matter density parameter $\Omega_m$. A representative example is GW170817 \cite{LIGOScientific:2017vwq}, the first binary neutron star merger with EM counterparts, which yielded the first GW-based standard siren measurement of $H_0$ and demonstrated the potential of this method for cosmology \cite{LIGOScientific:2017adf}.
%Notably, they offer a promising approach to address the Hubble tension arising from the inconsistency between low-redshift ($z\lesssim2$) observations from type-Ia supernovae \cite{Riess:2021jrx} and high-redshift ($z\sim1000$) measurements from the cosmic microwave background \cite{Planck:2018vyg}.

%GW observations provide an entirely independent and self-calibrating probe of the cosmic expansion, which can help arbitrate between these conflicting measurements and shed light on possible new physics beyond the standard cosmological model.

GW standard sirens face several limitations in constraining cosmological parameters, primarily due to the mass-redshift degeneracy~\cite{Schutz:1986gp} (therefore lack of redshift information) and uncertainties in luminosity distance estimates \cite{Chassande-Mottin:2019nnz}. Moreover, their effectiveness traditionally relies on the accuracy of waveform templates and the precision of the matched-filtering technique, which must simultaneously infer more than ten waveform parameters.
One proposed alternative is to use the time delays of strongly lensed GWs, in combination with the redshifts and images provided by their EM counterparts such as kilonovae, short gamma-ray bursts, or fast radio bursts \cite{Liao:2017ioi}.
%One proposed alternative is to use time delays from strongly lensed GWs in combination with the redshifts and images of their electromagnetic (EM) counterparts \cite{Liao:2017ioi}. 
The significantly higher precision of time-delay measurements from lensed GWs, compared to those from lensed quasars, enables more accurate cosmological constraints. However, since the detection of EM counterparts typically requires a neutron star in the binary system, the predominantly observed BBH merger events \cite{KAGRA:2021vkt} generally do not qualify. To address this limitation, a population-level method has been proposed that avoids reliance on EM counterparts \cite{Jana:2022shb, Jana:2024uta}, by leveraging statistical features of lensed GW catalogs, such as the number of detected lensing events and the distribution of time delays.

The effectiveness of population-based approach relies on the detection of a large number of lensed GW events. The third generation GW detectors, such as the Einstein Telescope \cite{Punturo:2010zz} and Cosmic Explorer \cite{Reitze:2019iox}, are expected to achieve significantly enhanced sensitivity and broader frequency coverage. These detectors will be capable of detecting GW signals from stellar mass BBHs at redshifts of up to $z\sim100$ \cite{Hall:2019xmm}, recording millions of BBH events per year \cite{ET:2019dnz}, compared to only $\mathcal{O}(100)$ events observed by the LIGO–Virgo–KAGRA to date. This unprecedented event rate provides a solid foundation for implementing population-based statistical analyses.

%In particular, recent studies \cite{Jana:2022shb,Jana:2024uta} have shown that the population properties of strongly lensed GW events, such as the distribution of lensing time delays and the number of lensed events, can be used to constrain cosmological parameters, especially in the redshift regime $z\sim10$, which is typically not accessible by other probes.%, assuming prior knowledge of the distributions of the source redshifts and lens properties. 
% This method has been shown to provide precise constraints on parameters like $\{H_0,\Omega_m\}$ in $\Lambda$CDM cosmology and $\{w_0,\Omega_m\}$ in $w$CDM cosmology.

Beyond cosmology, gravitational lensing offers a unique opportunity to test the validity of general relativity (GR) at galaxy scales. The parameterized post-Newtonian (PPN) framework \cite{Thorne:1971iat}, particularly the parameter $\gamma$, quantifies deviations from GR. While precise constraints on $\gamma$ have been obtained through solar system experiments -- such as the Cassini mission \cite{Bertotti:2003rm} -- constraints at galactic and cosmological scales remain relatively unexplored. 
%\textcolor{blue}{At these scales, $\gamma$ can also lead to degeneracies with cosmological parameters in lensing-based inference.}
Currently, most galaxy-scale constraints on $\gamma$ depend on precisely resolved stellar kinematics of the lens galaxy and prior knowledge of cosmological parameters, as demonstrated in observations of lenses such as ESO 325-G004 \cite{Collett:2018gpf}, RXJ1121-1231, and B1608+656 \cite{Jyoti:2019pez}. Only a limited number of studies have jointly constrained the Hubble constant and the PPN parameter $\gamma$ using lensed EM signals, including those from quasars \cite{Yang:2020eoh, Liu:2023ulr, Gao:2022ifq}. Table~\ref{tab:constraintsongamma} lists several constraints of $\gamma$ at different scales.

\begin{table}[t]
	\centering
	\renewcommand{\arraystretch}{1.5}
	\begin{tabularx}{\columnwidth}{>{\centering\arraybackslash}X >{\centering\arraybackslash}X >{\centering\arraybackslash}X} 
		\hline\hline
		Scale & Constraint on $\gamma$ & Reference\\ 
		\hline
		Solar system & $ 1 + (2.1 \pm 2.3) \times 10^{-5}$ & Ref. \cite{Bertotti:2003rm}         \\
		\hline
		Galaxy &  $0.97\pm0.09$ & Ref. \cite{Collett:2018gpf}         \\ 
		\hline
		Galaxy & $0.87^{+0.19}_{-0.17}$ & Ref. \cite{Yang:2020eoh} \\
		\hline
	\end{tabularx}
	\caption{Representative constraints on $\gamma$ at different scales. % at $68\%$ credible level
	}
	\label{tab:constraintsongamma}
\end{table}

On the other hand, lensed GWs offer significant advantages over traditional lensed EM sources such as quasars, including higher precision in time-delay measurements (the arrival times can be measured with precision of $\sim 10^{-4}$ s) and a substantially higher detection rate. While quasars are typically confined to low redshifts (e.g., $z \sim 1$), GW lensing can probe a much broader redshift range, extending up to $z \sim 10$. These unique characteristics make lensed GWs a promising tool for jointly constraining cosmological parameters and testing gravity, reaching beyond the capabilities of current observational methods.

While unlensed GWs can also provide valuable tests of GR through waveform-parameterized method or propagation of GW \cite{LIGOScientific:2019fpa, LIGOScientific:2016lio, Will:2014kxa, Belgacem:2017ihm, Yang:2021qge}, their sensitivity and accuracy are highly dependent on the precision of waveform modeling and parameter estimation. In contrast, the population-based lensed GW approach does not rely on the waveform of the signals. Given the low uncertainties in both time-delay measurements and the count of lensed events, this method is expected to serve as a more powerful and efficient probe of gravity, offering lower statistical error and significantly reduced computational cost.

In this paper, for the first time, we propose a unified Bayesian framework that jointly constrains the Hubble constant and the post-Newtonian parameter using only the population properties -- time-delay distributions and event counts -- of lensed GW events. This approach is independent of electromagnetic counterparts, waveform modeling, and resolved stellar kinematics of the lens galaxy.
%By combining the time-delay distribution with the number of lensed events, we quantify the joint sensitivity to both the background cosmology and potential deviations from general relativity. 
Assuming a flat $\Lambda$CDM cosmology, we parametrize the Hubble constant as $H_0 = 100h~\text{km}\text{s}^{-1}\text{Mpc}^{-1}$. Our results show that, under two different assumptions for the prior of $\Omega_m$, the uncertainties in $h$ and $\gamma$ fall within the ranges of $0.60\% \sim 0.99\%$ and $0.53\%\sim 3.3\%$, respectively, at the $68\%$ credible level -- representing a significant improvement over existing joint constraints.

\section{Methods}

We focus on the gravity test at the lensing galaxy scales and model the modified gravity effect via PPN framework with a constant parameter $\gamma$. 
%Since this parametrization cannot be extended to the cosmological scales, the distance-redshift relation won't change.
The time delay between lensed images $A$ and $B$ is given by 
\begin{align}
    \Delta t_{AB}=(1+z_l)\frac{D_s D_l}{c D_{ls}}\left[\phi(\theta_A,\beta)-\phi(\theta_B,\beta)\right],\label{eq:deftimedelay}
\end{align}
where $z_l$ is the redshift of lens, $\phi(\theta,\beta)=(\theta-\beta)^2/2-\psi(\theta)$ is called the Fermat potential, and $D_s$, $D_l$ and $D_{ls}$ are the angular diameter distances from the earth to the source, from the earth to the lens and from the lens to the source respectively. 
We model lens galaxies with a singular isothermal sphere (SIS), adopted as an effective description of the lens population at the statistical level. While individual lenses can deviate from a perfect isothermal profile due to ellipticity, external shear, or more complex mass distributions, observational studies (e.g., SLACS) show that these deviations, quantified in terms of the fitted logarithmic slope of the total mass density profile, differ from the isothermal value by only $\sim 3.7\%$ in the mean, with an intrinsic scatter of order $\sim 7\%$, and these deviations are predominantly stochastic across the population, leading to statistical cancellation at the population level \cite{Gavazzi:2007vw, Koopmans:2009av, Barnabe:2011gb}. As a result, the ensemble-averaged lensing properties—particularly those relevant for time-delay statistics—are well captured by an effective isothermal description. Under the SIS model, strong lensing of a point source generically produces double-image systems, which we adopt as the fiducial configuration. Then the lensing potential is $\psi(\theta)=\theta_E|\theta|$, where $\theta_E$ is the angular Einstein radius. The images' positions $\theta_{A,B}$ are solved from the lensing equation $\beta=\theta-\alpha(\theta)$ \cite{Schneider:1992bmb}. 
The deflection angle $\alpha(\theta)$ is obtained by taking the gradient of the lensing potential: $\alpha(\theta)=\nabla\psi(\theta)= \mathrm{sgn}(\theta) \theta_E$. 

In the PPN framework, the lensing potential is rescaled as $\psi^{\text{PPN}} = \frac{1+\gamma}{2} \psi^{\text{GR}}$ \cite{will2018theory} and consequently, the deflection angle is modified to $\alpha^{\text{PPN}}=\frac{1+\gamma}{2}\alpha^{\text{GR}}$. From the lensing equation, the image positions can be solved as $\theta_{\pm}=\beta\pm\theta_E^{\text{PPN}}$,
where $\theta_E^{\text{PPN}} = \frac{1+\gamma}{2}\theta_E^{\text{GR}}$ \cite{Bolton:2006yz}.
Substituting these results into the time delay expression Eq. \eqref{eq:deftimedelay}, 
one can find that the time delay in the PPN framework becomes $\Delta t^{\text{PPN}}=\frac{1+\gamma}{2} \Delta t^{\text{GR}}$.
With the expression of $\Delta t^{\text{GR}}$ in Ref. \cite{Oguri:2001qu}, the time delay with PPN correction is given by
\begin{align}
	\Delta t^{\text{PPN}} = 32\pi^2\frac{1+\gamma}{2}\frac{y}{c}\left(\frac{\sigma}{c}\right)^4  (1+z_l) \frac{D_{ls}D_l}{D_{s}},
\end{align}
where $y$ is the impact parameter, $\sigma$ is the lens velocity dispersion and $z_s$ is the redshift of source. 

We model the lens as a SIS with a modified Schechter function \cite{SDSS:2003fyb}. The differential optical depth in GR is given by \cite{Sereno:2011ty}
\begin{align}
	\frac{\partial^2\tau}{\partial z_l\partial\sigma}
	=\frac{\partial n}{\partial \sigma}(z_l,\sigma)s_{cr}(z_l,\sigma)c\frac{\mathrm{d}t}{\mathrm{d}z_l}(z_l),\label{eq:DifferentialOpticalDepth}
\end{align}
where $\frac{\partial n}{\partial \sigma}$ is modified Schechter function,
\begin{align}
	&\frac{\partial n}{\partial \sigma}(z_l,\sigma)=\frac{n_*}{\sigma_*} \frac{\beta}{\Gamma(\alpha/\beta)} \left(\frac{\sigma}{\sigma_*}\right)^{\alpha-1} \mathrm{exp}\left[-\left(\frac{\sigma}{\sigma_*}\right)^\beta\right] \,,
\end{align}
%According to the definition of redshift, we know that  $\mathrm{d}t/\mathrm{d}z_l=-1/\left[(1+z_l) H_0 E(z_l)\right]$. 
and $s_{cr}$ is the cross section \cite{Sereno:2011ty}. 
After introducing the PPN correction, it becomes
\begin{align}
	s_{cr}^{\text{PPN}}=\pi D_l^2\left(\theta_E^{\text{PPN}} \right)^2\left[y_\text{max}^2-\frac{2\Delta t^{\text{PPN}}}{3 T_{\text{obs}} y}y_\text{max}^3\right], 
\end{align}
where the $y_{\text{max}}$ is the maximum impact parameter and the $T_{\text{obs}}$ is the observation duration. Integrating Eq. \eqref{eq:DifferentialOpticalDepth} generates the optical depth with PPN correction.
%\begin{align}
%	\tau(z_s,\boldsymbol{\Omega})
%	=&\left(\frac{1+\gamma}{2}\right)^2\frac{F_*}{30}\left[\left(1+z_s\right)D_s\right]^3 y_\text{max}^2\notag\\
%    &\times\left[1-\frac{1+\gamma}{2}\frac{\Delta t_*}{7T_\text{obs}}\frac{\Gamma\left[(8+\alpha)/\beta\right]}{\Gamma\left[(4+\alpha)/\beta\right]}\right],
%\end{align}
%where
%\begin{align}
%	F_*&=16\pi^3 n_*\frac{\sigma_*^4}{c^4}\frac{\Gamma\left[(4+\alpha)/\beta\right]}{\Gamma(\alpha/\beta)},\\
%	\Delta t_*&=32\pi^2 \frac{\sigma_*^4}{c^4} \frac{D_s}{c} (1+z_s)y_{\text{max}}.
%\end{align}
Here the parameters $\{n_*, \alpha, \beta, \sigma_*\}$, called the velocity dispersion function parameters, have been extensively measured through observations of diverse galaxy populations \cite{SDSS:2003fyb, Montero_Dorta_2017, Choi:2006qg, Chae:2008ni, Mitchell:2004gw}.

Now, assuming $N$ lensed events have been detected with exact time delays $\{\Delta t_i\}_{i=1}^{N}$ between each lensed binary image signals during the observation period $T_\text{obs}$, the population characteristics of lensed GWs can be determined by two independent likelihoods given by $N$ and $\{\Delta t_i\}$ respectively \cite{Jana:2022shb}:
\begin{align}
\mathcal{L}\left(N,\{\Delta t_i\}|\boldsymbol{\Omega},T_{\text{obs}}\right)=\mathcal{L}\left(N|\boldsymbol{\Omega},T_{\text{obs}}\right)\times\mathcal{L}\left(\{\Delta t_i\}|\boldsymbol{\Omega},T_{\text{obs}}\right),
\end{align}
where $\boldsymbol{\Omega}=(h,\Omega_m,\gamma)$.

Under the assumptions of event independence and negligible probability for multiple events occurring in a short interval, the likelihood  of observing $N$ lensed events is described by a Poisson distribution:
\begin{align}
	\mathcal{L}\left(N|\boldsymbol{\Omega},T_{\text{obs}}\right)=\frac{\Lambda(\boldsymbol{\Omega},T_{\text{obs}})^N e^{-\Lambda(\boldsymbol{\Omega},T_{\text{obs}})}}{N!}.\label{eq:combinedlikelihood}
\end{align}
Here $\Lambda(\boldsymbol{\Omega},T_{\text{obs}})$ is the expected number of lensed events over the observation duration $T_{\text{obs}}$, which can be expressed as \cite{Jana:2022shb}
\begin{align}
	\Lambda(\boldsymbol{\Omega},T_{\text{obs}})=&R\int_{0}^{z_{\text{max}}}p_b(z_s|\boldsymbol{\Omega})P_l(z_s|\boldsymbol{\Omega})\mathrm{d}z_s\notag\\
    &\times\int_{0}^{T_{\text{obs}}}p(\Delta t|\boldsymbol{\Omega})(T_{\text{obs}}-\Delta t)\mathrm{d}\Delta t,\label{eq:Lambda}
\end{align}
where $R$ denotes the number of BBH mergers detectable per year. The functions $p_b(z_s|\boldsymbol{\Omega})$, $P_l(z_s|\boldsymbol{\Omega})$ and $p(\Delta t|\boldsymbol{\Omega})$ are the redshift distribution of the merging binaries sources, the strong lensing probability, and the time delay distribution respectively. It's worth noting that the maximum source redshift $z_\text{max}$ needs to be rescaled for different values of $h$ and $\Omega_m$, since the same detector sensitivity limit corresponds to the same maximum detectable luminosity distance. 
%In the context of flat $\Lambda$CDM cosmology, the luminosity distance is given by
%\begin{align}
%	d_L(z)=(1+z)\frac{c}{H_0}\int_{0}^{z}\frac{\mathrm{d}z'}{E(z')}, %which is an elliptic integral.
%\end{align}
%where $E(z)=\sqrt{\Omega_m(1+z)^3+\Omega_\Lambda}$ with $\Omega_\Lambda=1-\Omega_m$.

Considering the independence among different events, the likelihood of observing the time delays $\{\Delta t_i\}$ is given by
\begin{align}
	p\left(\{\Delta t_i\}|\boldsymbol{\Omega},T_{\text{obs}}\right)=\prod_{i=1}^{N}p(\Delta t_i|\boldsymbol{\Omega},T_{\text{obs}}),
\end{align}
where $p(\Delta t_i|\boldsymbol{\Omega},T_{\text{obs}})$ is the value of the model time delay distribution $p(\Delta t|\boldsymbol{\Omega},T_{\text{obs}})$ evaluated at $\Delta t_i$.  The model time delay distribution is given by
\begin{align}
	p(\Delta t|\boldsymbol{\Omega},T_{\text{obs}})\propto p(\Delta t|\boldsymbol{\Omega}) (T_{\text{obs}}-\Delta t)\Theta(T_{\text{obs}}-\Delta t),\label{eq:observed_pdt}
\end{align}
where $\Theta$ denotes the Heaviside step function.

To construct the redshift distribution of GW sources $p_b(z_s|\boldsymbol{\Omega})$, we assume the merger rate is uniform in comoving volume. The differential comoving volume at $z_s$ is $\dot{V_c}(z_s)=4\pi c(1+z_s)^2 D_s^2/\left[H_0 E(z_s)\right]$. The observed merger rate includes a time dilation factor $(1+z_s)^{-1}$, leading to $p_b(z_s|\boldsymbol{\Omega})\propto \dot{V_c}(z_s) /(1+z_s)$.

Considering the low proportion of lensed events among all GW events as suggested by current detections \cite{LIGOScientific:2021izm}, the strong lensing probability can be approximated to be  $P_l(z_s|\boldsymbol{\Omega})=1-e^{-\tau(z_s,\boldsymbol{\Omega})}\approx\tau(z_s,\boldsymbol{\Omega})$, where $\tau(z_s,\boldsymbol{\Omega})$ is the strong lensing optical depth for an event originating from a source at redshift $z_s$.

The expected time delay distribution $p(\Delta t|\boldsymbol{\Omega})$ is generated by marginalizing the parameters $\boldsymbol{\lambda}=(y,\sigma,z_l,z_s)$, i.e.,
\begin{align}
	p(\Delta t|\boldsymbol{\Omega})=\int p(\Delta t|\boldsymbol{\lambda},\boldsymbol{\Omega})p(\boldsymbol{\lambda}|\boldsymbol{\Omega})\mathrm{d}\boldsymbol{\lambda},\label{eq:pdt}
\end{align} 
where $p(\Delta t|\boldsymbol{\lambda},\boldsymbol{\Omega})$ is the distribution of the time delay $\Delta t$ given $\boldsymbol{\lambda}$ and $\boldsymbol{\Omega}$, and $p(\boldsymbol{\lambda}|\boldsymbol{\Omega})$ is the probability distribution of the parameter vector $\boldsymbol{\lambda}$ given $\boldsymbol{\Omega}$. The latter can be split as 
\begin{align}
    p(\boldsymbol{\lambda}|\boldsymbol{\Omega})=p(\sigma,z_l|z_s,\boldsymbol{\Omega})p_b(z_s|\boldsymbol{\Omega})p(y|\boldsymbol{\Omega}),
\end{align}
where $p(\sigma,z_l|z_s,\boldsymbol{\Omega})\propto \partial^2\tau/\partial z_l\partial\sigma$. The distribution of lensed sources is assumed to be uniform within the Einstein radius on the lens plane, resulting in $p(y) \propto y$ for $y\in [0,1]$.

Given the parameters $\boldsymbol{\lambda}$ and $\boldsymbol{\Omega}$, the value of $\Delta t$ is uniquely determined, resulting in its probability distribution being a Dirac delta function centered at the calculated value: $p(\Delta t|\boldsymbol{\lambda},\boldsymbol{\Omega})=\delta(\Delta t - \Delta t(\boldsymbol{\lambda},\boldsymbol{\Omega}))$.
%The optical depth $ \tau $ characterizes the probability that a light ray is lensed when propagating from the source (redshift $ z_s $) to the observer. For a given $ z_s $, the total optical depth $ \tau(z_s, \boldsymbol{\Omega}) $ is approximately the probability of a lensing event (when $ \tau \ll 1 $). 
We also get $p(\sigma,z_l|z_s,\boldsymbol{\Omega})\propto \partial^2\tau/\partial z_l\partial\sigma$. The reason is as follows. The differential optical depth describes the contribution to the total optical depth in the unit interval at the lens redshift $ z_l $ and velocity dispersion $ \sigma $, i.e., the probability density of the lensing event for this combination of parameters. The conditional probability $ p(\sigma, z_l | z_s, \boldsymbol{\Omega}) $ represents the joint distribution of the lens parameters $ (\sigma, z_l) $ given the source redshift $ z_s $ and $\boldsymbol{\Omega}$. The differential optical depth directly reflects the contribution weight of lenses with different $ z_l $ and $ \sigma $ to the total optical depth, so it naturally becomes the physical basis of this probability distribution.

%After normalization, $p(y|\boldsymbol{\Omega})=2y$. However, the actual distributions can be inferred from the statistics of lensed signals' magnification from the following relationship in the SIS model: $\mu_\pm=(1/y)\pm1$.

\section{Simulations and results}

We assume the maximum of detectable redshift of the source under the “true” cosmology $(h=0.7,\Omega_m=0.3, \gamma=1)$ is $z_{\text{max}}=20$, consistent with the expected horizon of third-generation detectors \cite{Reitze:2019iox}. The lensing galaxy model is based on the velocity dispersion function parameter measurements reported in Ref. \cite{Montero_Dorta_2017}. The default merger rate and observation duration are $R=5\times10^5\text{yr}^{-1}$ and $T_\text{obs}=10\text{yrs}$ respectively, consistent with projected capabilities discussed in Refs. \cite{Punturo:2010zz, Reitze:2019iox, ET:2019dnz}.

%Notably, the integral in Eq. \eqref{eq:pdt} typically does not have an analytical solution, so we will adopt Markov Chain Monte Carlo (MCMC) from the Python package {\sc emcee} \cite{Foreman-Mackey:2012any} to generate samples and compute the distribution via statistical sampling. 

\begin{figure*}[htbp]
	\includegraphics[width=0.32\textwidth]{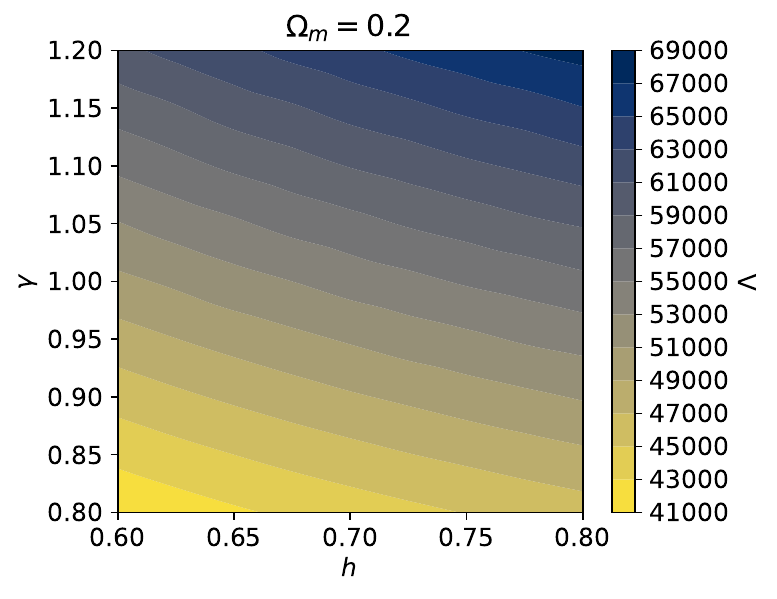} 
	\includegraphics[width=0.32\textwidth]{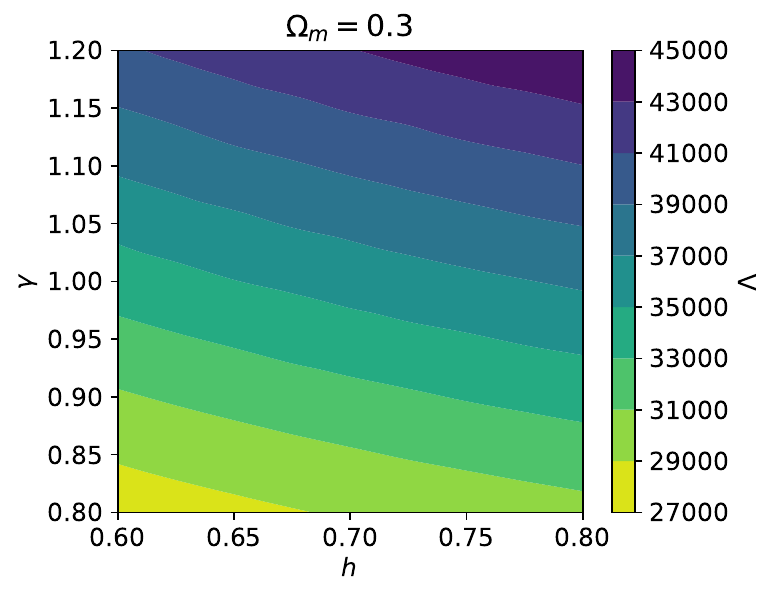} 
	\includegraphics[width=0.32\textwidth]{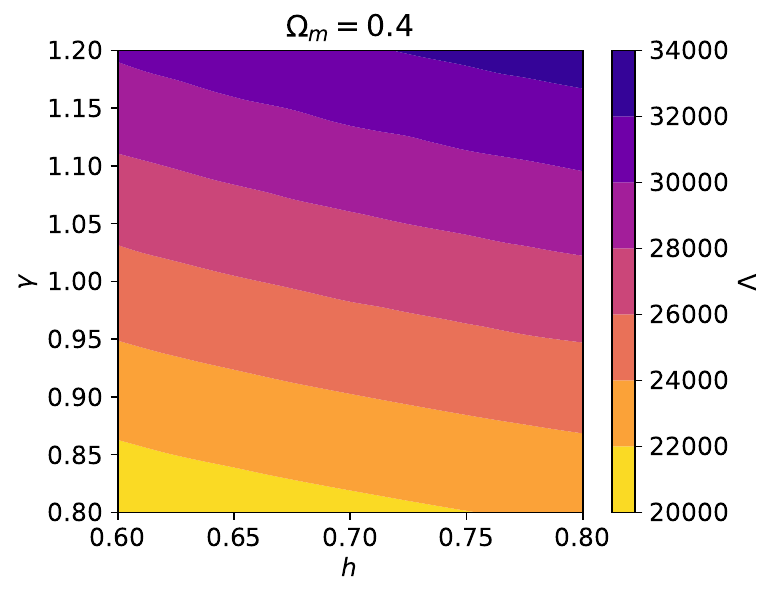} 
	\caption{Distributions of the lensed event number $\Lambda$ in the $\gamma-h$ parameter space for three fixed values of $\Omega_m$. }
	\label{fig:Lambda_om}
\end{figure*}

To explore the dependence of the expected number of lensed events $\Lambda$ on $\boldsymbol{\Omega}$, we present comparative plots in Fig. \ref{fig:Lambda_om}. We find that the same value of $\Lambda$ can map to different combinations of $(h, \Omega_m, \gamma)$, indicating degeneracies among these parameters. These parameter dependencies determine the degeneracy structure and constraining power in the Bayesian inference presented below. Fig. \ref{fig:Lambda_om} also illustrates that larger values of $h$ and $\gamma$ tend to increase the number of lensed events, while increasing $\Omega_m$ has the opposite effect. This trend can be understood as a consequence of the geometric dependence of angular diameter distances on $\Omega_m$ in our fixed-lens-population model, which results in a suppression of the lensing rate for larger $\Omega_m$. This behavior is consistent with Ref.~\cite{Cooray:1999ru}, but differs from Ref.~\cite{Jana:2022shb}, where higher $\Omega_m$ leads to more lensed events. In Ref.~\cite{Jana:2022shb}, $\Omega_m$ affects both cosmological distances and the lens population, with larger values resulting in more massive halos, thereby introducing additional sensitivity to $\Omega_m$ beyond geometric effects.
%However, we note that Ref. \cite{Jana:2022shb} reports an opposite trend, where a higher $\Omega_m$ leads to more lensed events. The discrepancy arises because, in Ref. \cite{Jana:2022shb}, $\Omega_m$ not only affects the cosmological distances but also plays a direct role in the construction of the lens population. Specifically, in that simulation framework, the mass distribution of lenses is explicitly dependent on $\Omega_m$, with larger $\Omega_m$ leading to more massive lensing halos. This introduces an additional sensitivity to $\Omega_m$ beyond the pure geometric effects, significantly influencing the predicted number of lensed events.

\begin{figure}[htbp]
\raggedright
\includegraphics[width=0.5\textwidth]{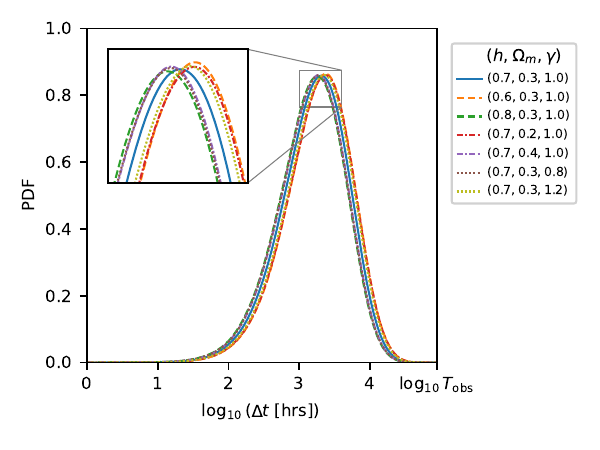} 
\caption{Observed PDF of the time delays for the fiducial parameter set $\boldsymbol{\Omega}=(0.7,0.3,1.0)$ and one-parameter variations around it.}
\label{fig:pdtMD}
\end{figure}

We then examine the influence of $\boldsymbol{\Omega}$ on the distribution of observed time delays $p(\Delta t|\boldsymbol{\Omega},T_\text{obs})$. As shown in Fig.~\ref{fig:pdtMD}, variations of $\boldsymbol{\Omega}$ result in shifts in the shape and peak location of the time-delay PDF. Specifically, increasing $h$ or $\Omega_m$ shifts the peak of the distribution toward shorter time delays, while increasing $\gamma$ moves the peak toward longer time delays. Besides, the overlap of different lines suggests the occurrence of degeneracy among $h$, $\Omega_m$ and $\gamma$. These shifts and degeneracies directly determine the structure of the constraints obtained from the time-delay population analysis. The concentration of time delays around several months ($\sim 10^3$ hours, i.e., $\log_{10}(\Delta t\mathrm{[hrs]})\sim 3)$ aligns with expectations for galaxy-scale gravitational lensing of GW \cite{Oguri:2018muv,LIGOScientific:2021izm}. The shifting behavior with respect to $\Omega_m$ is qualitatively consistent with trends reported in Ref. \cite{Harvey:2020lwf,Wei:2017emo}. These shifts arise from the dependence of the time delay in Eq.~\eqref{eq:deftimedelay}, which propagates into the PDF $p(\Delta t|\boldsymbol{\Omega})$. This distribution then determines the expected number of detected events via Eq.~\eqref{eq:Lambda} and enters the likelihood of the observed time delays in Eq.~\eqref{eq:observed_pdt}, thereby directly shaping the constraints on $\boldsymbol{\Omega}$.

\begin{figure}[htbp]
\includegraphics[width=0.5\textwidth]{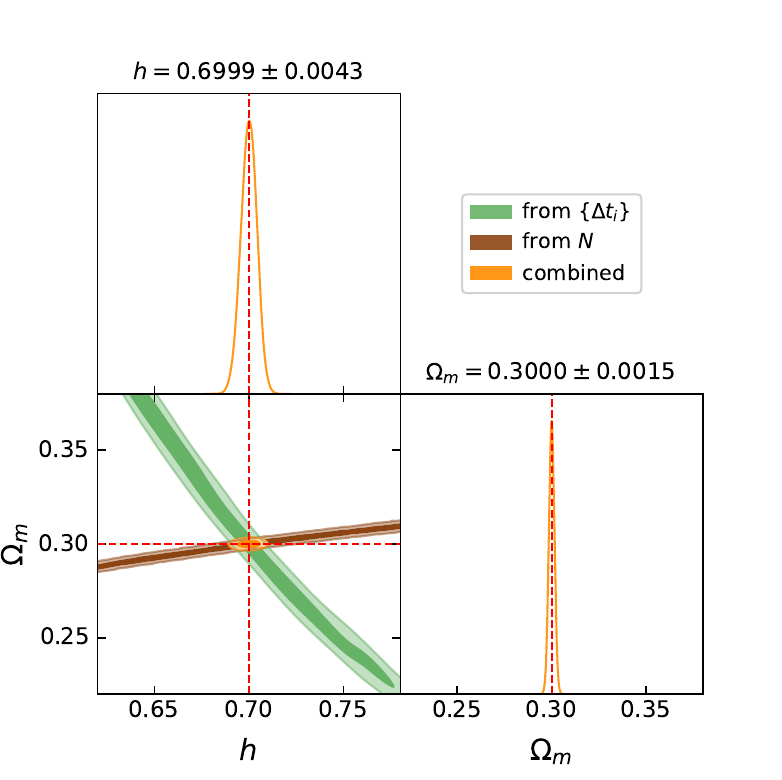} 
\caption{Posterior distributions ($68\%$ and $95\%$ credible regions) of $h$ and $\Omega_m$ with $\gamma$ fixed at $1$. The $68\%$ credible intervals yield the constraints: $h=0.6999\pm0.0043$ and $\Omega_m=0.3000\pm0.0015$.}
\label{fig:hom}
\end{figure}

%To study the parameter constraint results, we adopt uniform priors over the following ranges: $0.6<h<0.8,\,0.2<\Omega_m<0.4,\,0.8<\gamma<1.2$. 
According to Bayes' theorem, the posterior is proportional to the product of the likelihood Eq.~\eqref{eq:combinedlikelihood} and the prior. We compute the expected number of lensed events $\Lambda$ using Eq.~\eqref{eq:Lambda}, and draw the simulated observed number $N$ from a Poisson distribution with mean $\Lambda$, with the corresponding time delays $\{\Delta t_i\}$ from Eq.~\eqref{eq:observed_pdt}, assuming the true parameters $(h, \Omega_m, \gamma) = (0.7, 0.3, 1)$.
%We use Eqs. \eqref{eq:Lambda} and \eqref{eq:observed_pdt} to simulate the observed event number $N$ and time delays $\{\Delta t_i\}$ respectively, assuming the true parameters $(h,\,\Omega_m,\,\gamma)=(0.7,\,0.3,\,1)$. 
We begin by constraining the cosmological parameters in GR. Fig. \ref{fig:hom} presents the joint constraints on $h$ and $\Omega_m$. Uniform priors over the following ranges are adopted: $0.6<h<0.8,\,0.2<\Omega_m<0.4$, with $\gamma$ fixed at $1$. The uncertainties are found to be $0.43\%$ for $h$ and $0.15\%$ for $\Omega_m$, compared to $1.4\%$ for $h$ from SH0ES \cite{Riess:2021jrx}, and $0.7\%$ for $h$ and $2.2\%$ for $\Omega_m$ from Planck \cite{Planck:2018vyg}, assuming GR holds.

%It is evident that our constraints of $h$ and $\Omega_m$ are tighter compared to those reported in Ref. \cite{Jana:2022shb}. This improvement arises from the nature of our time delay distribution. In our model, the PDF $p(\Delta t|\boldsymbol{\Omega})$ naturally approaches zero before reaching the observational time limit $T_\text{obs}$. As a result, the truncation at $T_\text{obs}$ does not lead to a loss of information; that is, $p(\Delta t|\boldsymbol{\Omega})\approx p(\Delta t|\boldsymbol{\Omega},T_\text{obs})$. This ensures that the full range of time delays contributes effectively to the likelihood. In contrast, in Ref. \cite{Jana:2022shb}, the PDF $p(\Delta t|\boldsymbol{\Omega})$ does not vanish at $T_\text{obs}$. Then, the truncation procedure reduces the sensitivity of the likelihood to time delays near $T_\text{obs}$ (see the right panel of Figure 1 in Ref. \cite{Jana:2022shb}), thereby leading to a loss of information and broader credible regions.

\begin{figure}[htbp]
\includegraphics[width=0.5\textwidth]{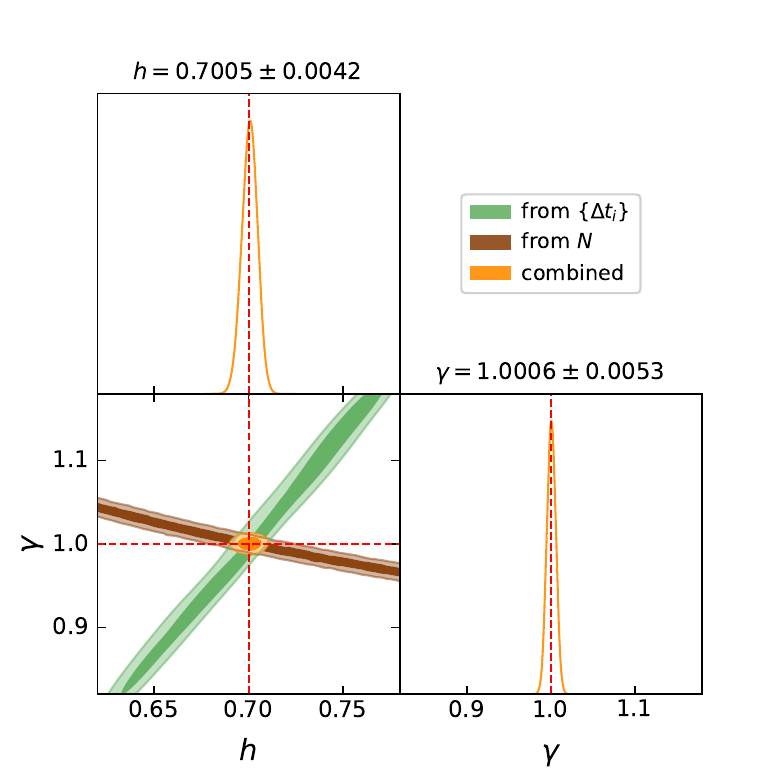} 
\caption{Posterior distributions ($68\%$ and $95\%$ credible regions) of $h$ and $\gamma$ with $\Omega_m$ fixed at $0.3$. The $68\%$ credible intervals yield the constraints: $h=0.7005\pm0.0042$ and $\gamma=1.0006\pm0.0053$.}
\label{fig:hppn}
\end{figure}

\begin{figure}[htbp]
\includegraphics[width=0.5\textwidth]{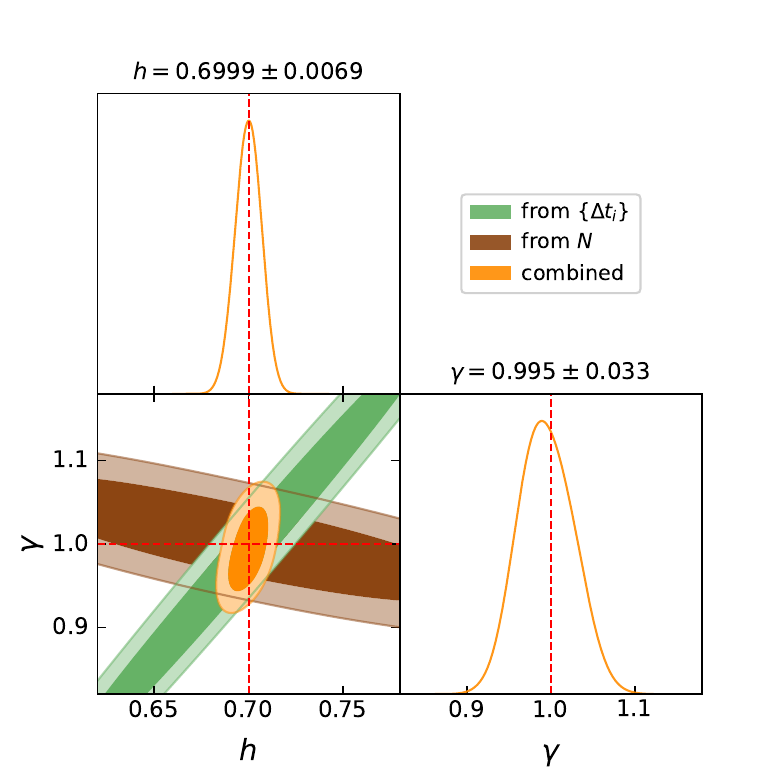} 
\caption{Posterior distributions ($68\%$ and $95\%$ credible regions) of $h$ and $\gamma$ with $\Omega_m$ assumed to follow a Gaussian distribution with the standard deviation $\sigma=0.0056$. The $68\%$ credible intervals yield the constraints: $h=0.6999\pm0.0069$ and $\gamma=0.995\pm0.033$.}
\label{fig:hppngauss}
\end{figure}

Due to the strong degeneracy among the parameters ${h,\,\Omega_m,\,\gamma}$, and given our primary interest in constraining the cosmic expansion rate and testing gravity theories, we adopt two strategies to address this degeneracy while simultaneously constraining $h$ and $\gamma$. Uniform priors are applied within the ranges $0.6 < h < 0.8$ and $0.8 < \gamma < 1.2$.
First, we fix the matter density parameter at $\Omega_m = 0.3$. As shown in Fig. \ref{fig:hppn}, the degeneracy directions of $h$ and $\gamma$ from event counts and time-delay distributions are anti-aligned. This complementary behavior significantly helps break the degeneracy between the two parameters in the joint analysis, leading to tighter constraints. Under this setting, $h$ and $\gamma$ can be constrained to precisions of $0.60\%$ and $0.53\%$, respectively.
In a more realistic scenario, we treat $\Omega_m$ as a Gaussian-distributed parameter with a standard deviation of $\sigma = 0.0056$, informed by Planck 2018 results \cite{Planck:2018vyg}. As illustrated in Fig. \ref{fig:hppngauss}, this leads to slightly weaker constraints, with $h$ and $\gamma$ determined to within 0.99\% and 3.3\%, respectively.
In both cases, our method achieves significantly higher precision than the minimum uncertainties of 1.5\% for $h$ and 8.7\% for $\gamma$ reported by previous joint probes \cite{Yang:2020eoh,Liu:2023ulr,Gao:2022ifq}. Since a third independent probe with better than $2\%$ precision on $H_0$ is generally considered sufficient to arbitrate the Hubble tension \cite{Chen:2017rfc}, strongly lensed GW populations presented in our work are therefore capable of resolving it. These results demonstrate the strong potential of population-based statistics of lensed GW events for jointly constraining cosmic expansion and gravity theories.
%As expected, an increase in the uncertainty of $\Omega_m$ leads to a corresponding decrease in the precision of the constraints on $h$ and $\gamma$.

%In contrast to other galaxy-scale probes, the previous results are listed in Table \ref{tab:jointconstraint}. Among them, Ref. \cite{Yang:2020eoh} utilizes lensed quasar signals from four well-studied H0LiCOW systems. Ref. \cite{Liu:2023ulr} combines lensed quasar signals and unlensed Type Ia supernovae as independent distance indicators. Besides, Ref. \cite{Gao:2022ifq} investigates the potential of lensed fast radio bursts, demonstrating that their short duration and clean imaging could significantly enhance measurement precision. As can be seen, the constraints on $h$ and $\gamma$ inferred from  the population of lensed GWs  remain significantly more precise than previous results based on EM observations, even after accounting for the uncertainty of $\Omega_m$.

\begin{comment}

\begin{table}[h]
\centering
\renewcommand{\arraystretch}{1.5}
\begin{tabular}{c|c|c|c} 
\hline\hline
         & \multicolumn{1}{c|}{Ref. \cite{Yang:2020eoh} } 
         & \multicolumn{1}{c|}{Ref. \cite{Liu:2023ulr} }
         & \multicolumn{1}{c}{Ref. \cite{Gao:2022ifq} }                           \\ 
\hline
$~h~$ & $0.7365^{+0.0195}_{-0.0226}$ & $0.729^{+0.020}_{-0.023}$ & uncertainty $\sim 1.5\%$          \\
\hline
$~\gamma~$ & $0.87^{+0.19}_{0.17}$ & $0.89^{+0.17}_{-0.15}$ & uncertainty $\sim8.7\%$            \\ 
\hline
\end{tabular}
\caption{Previous simultaneous constraints of $h$ and $\gamma$ at $68\%$ credible level.}
\label{tab:jointconstraint}
\end{table}

\end{comment}

\section{Conclusion and discussion}

We investigate the joint constraints on the Hubble constant and the post-Newtonian parameter using the population statistics of strongly lensed gravitational waves from binary black hole mergers observed by third-generation detectors. 
This work is formulated as a simulation-level forecast for the next-generation (XG) GW era, aimed at assessing the achievable error bars on inferred parameters rather than providing a fully realistic treatment of present-day astrophysical uncertainties. In the XG era, millions of unlensed GW detections will tightly constrain the merger-rate history, while forthcoming wide and deep galaxy surveys such as LSST and Euclid will precisely characterize the lens population \cite{ferrami2025velocity, Geng:2021tiz}. Motivated by this observational landscape, we adopt several idealizations in this work. Specifically, we treat the merger rate and the lens-population model as known inputs and use simplified parametrizations to isolate the constraining power of lensed-event counts and time-delay distributions. We also adopt the SIS model to describe the average properties of the lens population in a statistical sense; consequently, only double-image systems are considered, while other image morphologies are left outside the scope of this study.

Our results (see Table~\ref{tab:priors_constraints}) show that this population-based approach achieves joint constraints of $0.60\%\sim0.99\%$ for $H_0$ and $0.53\%\sim3.3\%$ for $\gamma$, compared to $\sim2\%$ for $H_0$ and $\sim20\%$ for $\gamma$ from existing joint constraints, indicating significantly improved performance.
Importantly, the proposed method relies solely on the observed time-delay distribution and the total number of lensed GW events, without requiring electromagnetic counterparts, detailed GW waveform modeling, or resolved stellar kinematics of the lens galaxy. Due to the expected low uncertainties in both time-delay measurements and event counts, this approach holds strong potential for delivering precise and robust constraints. Lensed GW population statistics thus constitute a precise and robust probe for joint tests of cosmology and gravity.

\begin{table}[t!]
	\centering
	\resizebox{\columnwidth}{!}{%
		\begin{tabular}{lcccc}
			\hline\hline
			& Parameter & $\gamma = 1$ & $\Omega_m = 0.3$ & $\Omega_m \sim \mathcal{N}(0.3,\,0.0056)$ \\
			\hline
			\multirow{3}{*}{\textit{Priors}}
			& $h$ & $[0.6,\,0.8]$ & $[0.6,\,0.8]$ & $[0.6,\,0.8]$ \\
			& $\Omega_m$ & $[0.2,\,0.4]$ & fixed to $0.3$ & $\mathcal{N}(0.3,\,0.0056)$ \\
			& $\gamma$ & fixed to $1$ & $[0.8,\,1.2]$ & $[0.8,\,1.2]$ \\
			\hline
			\multirow{3}{*}{\textit{Posteriors}}
			& $h$ & $0.6999 \pm 0.0043$ & $0.7005 \pm 0.0042$ & $0.6999 \pm 0.0069$ \\
			& $\Omega_m$ & $0.3000 \pm 0.0015$ & -- & -- \\
			& $\gamma$ & -- & $1.0006 \pm 0.0053$ & $0.995 \pm 0.033$ \\
			\hline\hline
		\end{tabular}%
	}
	\caption{Summary of priors and posteriors in this work. All quoted uncertainties correspond to $68\%$ credible intervals.}
	\label{tab:priors_constraints}
\end{table}

%Several extensions merit future investigation. First, we will test alternative source–population models \cite{Dominik:2013tma,Belczynski:2016obo}; existing studies indicate that such choices mainly induce quantitative shifts in inferred parameters without qualitative changes \cite{Jana:2024uta}. Second, we will adopt more realistic lens models -- including the singular isothermal ellipsoid \cite{kormann1994isothermal}, generalized Navarro–Frenk–White \cite{Jing:1999ir}, and halo–mass–function–based prescriptions \cite{Behroozi:2012iw} -- and reassess how to embed the PPN parameter $\gamma$ in each framework. Third, we will extend the analysis beyond $\Lambda$CDM (e.g., $w$CDM) and consider constraints on additional PPN parameters. Finally, we will quantify systematic uncertainties, particularly those associated with the velocity–dispersion function, by propagating distributions of its parameters rather than fixing them. These developments will be reported in future work.

%\appendix
%\section{}

\begin{acknowledgments}
This work is supported by the National Natural Science Foundation of China Grants No. 12575063, and in part by ``the Special Funds for the Double First-Class Development of Wuhan University'' under the reference No. 2025-1302-010. The numerical calculations in this paper have been done on the supercomputing system in the Supercomputing Center of Wuhan University.
\end{acknowledgments}
\bibliography{ref}

\end{document}